\documentstyle[aps,epsfig]{revtex}
\begin{document}
\draft \twocolumn[\hsize\textwidth\columnwidth\hsize\csname
@twocolumnfalse\endcsname
\title{Time resolved nanosecond vortex dynamics in high Tc superconducting films }
\author{
Hern\'{a}n Ferrari$^1$
, David Ibaceta$^1$
, Victoria Bekeris$^1$
, Esteban Calzetta$^1$
and 
Luigi Correra$^2$}
\address{
$^1$Departamento de F\'{\i}sica, Universidad de Buenos Aires, Pabell\'{o}n I, Ciudad Universitaria, 1428 Buenos Aires, Argentina.\\
$^2$CNR-LAMEL, 40129 Bologna, Italy.
}
\date{\today{}}
\maketitle

\begin{abstract}
We report on the observation of the rearrangement of vortices in the ns time scale following a fast magnetic field variation due to selective local heating of a high \(T_{c}\) superconducting film.
Detailed simulations describing the measured photoinduced voltage signals along the film lead to a better understanding of vortex dynamics in fast regimes, where mean vortex velocities reach \(10^4\)~m/s. 
\end{abstract}
]

\narrowtext

Vortex dynamics in type II superconductors continues to be a stimulating research area. 
Striking memory effects and dynamical ordering of vortex matter in the low frequency domain have been observed recently \cite{nosprl}, and may lead in the future to switching devices \cite{switch}. 
However, the high frequency response of vortex matter is still an open question that needs to be addressed.

In the past, the use of conventional techniques has limited the observations of magnetic properties to the low frequency domain. 
These techniques have long intrinsic response and initial data-acquisition times, somewhere between fractions of seconds and minutes\cite{Yeshurum}. 
The study of short time vortex dynamics with novel non-conventional techniques is a requirement for the development of future  fast devices.

In this letter we report on the rearrangement of vortices in a high \(T_{c}\) superconducting sample over nanosecond time scales observed following a fast magnetic field variation due to partial heating.

\begin{figure}
\centerline{\epsfig{file=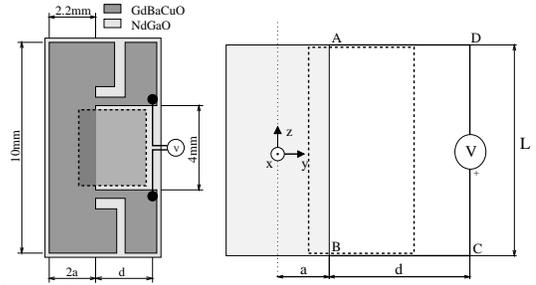,width=.80\hsize,clip=}}
\caption{Schematics of experimental geometry and simulations. The mobile laser spot is shown in dashed lines. The measure loop ABCD has dimensions \(L\) and \(d\).}
\label{Fig Schem}
\end{figure}

The sample is a GdBa$_2$Cu$_3$O$_{7-x}$ film of thickness \(\delta=300\)~nm on NdGaO$_3$ substrate.
The sample was patterned (see Fig. \ref{Fig Schem}) using laser ablation in a strip (\( \sim 10\times 2 \)~mm$^2$), and voltage contact pads were provided by gold sputtering, to reduce contact resistance below \(1~\Omega\).
The \(T_c\) of the film is \(91\)~K with a \(\Delta T_c=1.5\)~K (10\% - 90\% criterion) determined by \(ac\) susceptibility. 

For the experimental observations we used a high bandwidth electronic detection method coupled with a pulsed magnetic field generator and a synchronized pulsed laser, based on a previously designed short-time magnetometer \( ( \sim 200~\mu\)s ) \cite{nosrsi}.
The excimer laser provided \(45\)~ns FWHM pulses at \(308\)~nm and uniformly illuminated the target area.

Briefly, the technique consists in the preparation of the thermomagnetic initial state by stabilizing the sample temperature below the irreversibility line and applying a transverse pulsed magnetic field \(H_a\) at a high rate \( ( dH/dt\sim 1000\)~T/s) to reduce the uncertainty in the time origin. 
The laser pulse is triggered at a controlled delay, \( t_{d} \), at different stages of the propagation of the magnetic field front. 
Flux penetration can be followed, as will be described below, by measuring the voltage signal across the pair of contacts provided in the film. 
The optical heating lowers the pinning force on the vortex structure (at the location where the laser spot was positioned) below that corresponding to the critical state at uniform temperature. 
This allows the fast penetration of vortices and triggers the flux redistribution in the whole sample.
Flux motion is detected as a voltage pulse, superposed to signal induced by the application of the magnetic field.

It has been shown that the photoinduced voltage is related to the fast motion of vortices \cite{puig}. 
A thermal force cannot explain the results and instead a magnetic force drives flux motion.
Having control of the time delay and the location of the illuminated area of the film, it is possible to perform experiments that provide additional information on fast vortex dynamics \cite{Tesis,sst}.

The model in study can be understood from Fig. \ref{Fig Schem}. 
Assuming homogeneity in the \(z\) direction and uniform applied transverse magnetic field in the \(x\) direction, the voltage reading is given by \cite{Clem}:

\begin{equation}
\label{VoltageReading}
\frac{V}{L}=E_{b}+d\frac{dB_{a}}{dt}+\int ^{a+d}_{a}\frac{d\! B_{j}}{dt}dy
\end{equation}
where \( E_{b} \) is the electric field at the contacts edge of the strip, \( B_{a} \) is the applied induction field, \( B_{j} \) is the field produced by the current distribution, \(L\) and \(d\) are the dimensions of the ABCD loop. We are considering \(B=\mu_{0}H\).
The electric field \( E_{b} \) can be interpreted as given by the Josephson relation \cite{Joseph} \( E=-v\times B \), where \( v \) is an average vortex velocity and \( B \) the macroscopic magnetic induction.
The first term in Eq. (\ref{VoltageReading}) is a superposition of voltage pulses produced by individual vortices moving across the edge, each contributing in one flux quantum to the time integral of \( V(t) \).
The last term in Eq. (\ref{VoltageReading}) is the contribution from the change of the flux through the ABCD loop due to the return flux from the vortices.
A moving vortex that enters the sample from the left will contribute only to this term.

Assuming a phenomenological relationship between the electric field and the current density \(E=E(J,H,T)\), the basic features of the experimental results are well described by the quasi-stationary Maxwell equation \cite{brandt}:

\begin{equation}
\label{maxwell}
\frac{dE}{dy}=-\mu _{0}\left( \frac{dH_{a}}{dt}+\frac{\delta }{2\pi }\int ^{a}_{-a}\frac{d\! J}{dt}\frac{du}{u-y}\right)
\end{equation}

Our experiment samples both the high and low current response of the material.
In the high current regime above the critical state current density \(J_{c}\), the resistivity takes the flux flow value \( \rho_{ff}=\rho_{n} H/H_{c2} \), where \( H \) is the local field, \( H_{c2} \) is the upper critical one, and \(\rho_n\) is the normal state resistivity.
In the low current regime, \( J<J_{c} \), the resistivity is due to thermally activated creep and results in an electric field of the form \( E=E_{c}(J/J_{c})^{n} \) \cite{brandt}, where \( n>>1 \). 
Observe that the resistivity at each point depends on the local magnetic field, which in turn is sensitive to the current distribution over the whole sample, leading to a highly nonlinear and nonlocal dynamics.
The temperature dependence of the critical current is given by \(J_c(T)=J_{c0}(1-T/T_i)^{3/2}\) \cite{JcdeT}, where \(T_{i}=86\)~K is the observed irreversibility temperature above which no photo-signal is detected.
For the critical field, in the temperature range under study we can approximate 
\(H_{c2}(T)=(T_{c}-T)~1.8\times10^6\)~Am$^{-1}$K$^{-1}$ \cite{PFC}.

For the temperature independent parameters of these simulations we choose 
\( n=20 \), 
\( \delta J_{c0}=25\)~kA/m, 
\( \rho_{n}/\delta=100~\Omega\) and 
\(E_{c}=0.01\)~V/m.
All measurements (and calculations) were made at an initial temperature \(T_0=73.2\)~K.
The time-dependent temperature that results after local optical heating (see below) determines the time dependence of \(J_{c}\) and \(H_{c2}\).
 
To account for the experimental data we must assume that the crossover from one current regime to the other is smooth. 
We achieve this by interpolating both regimes with a cubic form in a small range \( (\sim 5\%~J_{c}) \).

Phenomenological constitutive relations of the kind we have discussed break down in the limit of very weak magnetic fields, i.e., when we are dealing with regions of the sample where the magnetic flux has not yet penetrated. 
For this reason, we assume that there is a cut-off in the constitutive relation, such that \(E=0\) whenever \(H<H_1\), with \(H_1=10^{-8}H_{c2}\). This regime is relevant to the early stages of the simulation.

The resulting nonlinear nonlocal magnetic flux diffusion equation was integrated numerically by means of a time adaptive algorithm for the generally non-symmetric sheet current distribution \(\delta J(y,t)\)\cite{Numerical Recipes}. 

\begin{figure}
\centerline{\epsfig{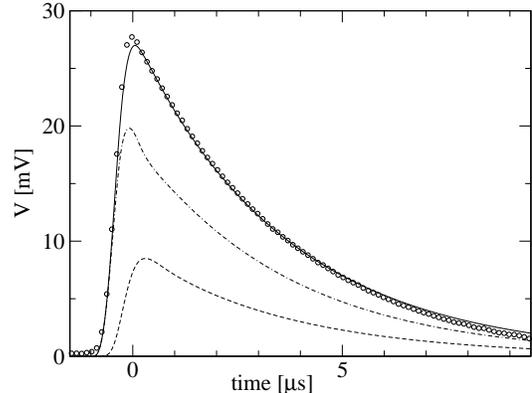}}
\caption{Observed and simulated signal for the turning-on of the applied field. From bottom to top, the lines represent the contribution of the electric field at the edge of the film, the contribution due to the flux variation into the ABCD loop, and the total signal. We define \(t=0\) at the maximum value of the measured signal.}
\label{Fig Turn}
\end{figure}

Figure \ref{Fig Turn} shows the voltage signal induced by the application of \( 7500\)~A/m.
The rising edge was approximately exponential with a characteristic rise time \( \tau \sim 3.5~\mu\)s. 
Symbols indicate the measured voltage, and in full line we show the calculated voltage signal with an applied field modeled as an algebraic initial slope followed by an exponential rising edge.
The lower signals, from bottom to top, correspond to the voltage on the edge of the sample, and the difference with the full signal.
Observe the delay in the onset of the edge electric field, associated with the formation of the characteristic current profile. 
Once this profile is formed, the model predicts the contacts voltage to be proportional to \( {dH_{a}}/{dt} \) and independent of the sample parameters for the regime where the applied field changes faster than the induced one. 

\begin{figure}
\centerline{\epsfig{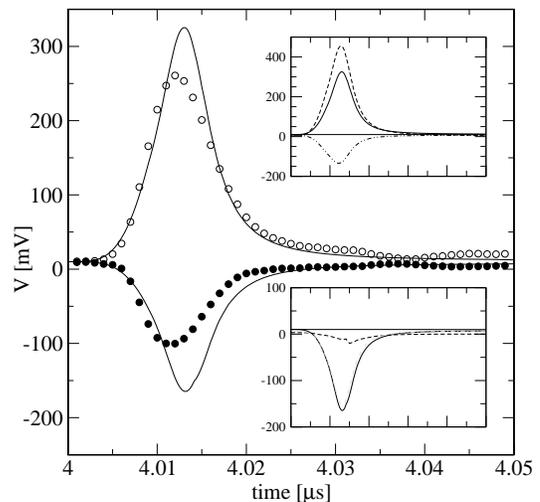}}
\caption{Voltage signals after optical heating of the sample. Open circles correspond to the illumination at the contacts side, and filled circles to illumination at the opposite one. Full lines show the simulated signals. The insets show the composition of these signals (see text).}
\label{Fig Sign}
\end{figure}

Illumination of the two thirds of the sample surface on the side of the contacts (right), \(4~\mu\)s  after the application of the field, produced the voltage pulse plotted in Fig. \ref{Fig Sign} (open circles); note the different scales.
The local temperature rise reduces the local critical current density below the flowing current density value (we neglect lateral thermal diffusion that would represent an increase in temperature in an additional small percentage of the non illuminated area during the duration of the pulse).
This leads to a rapid reduction of sheet current density and the correlated penetration of magnetic flux.

The curves plotted in the upper inset of the figure are the calculated voltage signals for this situation.
The dashed curve corresponds to the electric field contribution at the sample edge, the negative dotted curve is the term related to the magnetic flux variation through the ABCD loop, and the full line is the total calculated voltage signal also shown in the main panel.

\begin{figure}
\centerline{\epsfig{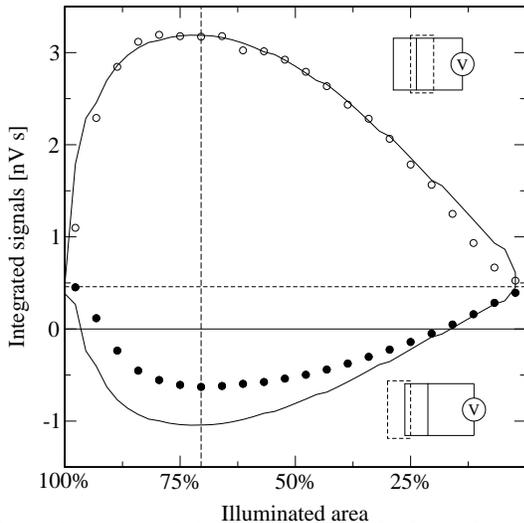}}
\caption{Scanning of the film, varying the heated area. The circles show the time integrated signals. Open (filled) circles correspond to heating from the contacts (opposite to contacts) side. The full lines show the simulated results. The horizontal dashed line shows the value obtained without heating. The vertical line shows the heated area chosen for other figures.}
\label{Fig Eyes}
\end{figure}

The filled circles show the corresponding result when a similar sample surface is illuminated at the edge opposite to the contacts (the left one). 
Notice that the measured voltage changes sign.
This result is well described by the calculations (lower inset), where the flux time derivative is almost the only contribution to the total voltage .
Indeed, almost no vortices move across the contacts edge as they are magnetically forced to enter the film from the opposite direction.

To test the model further, the laser spot was scanned across the film, repeating the experiment at a fixed temperature and at a fixed delay time \( t_{d} \).
The measured voltage signals were time integrated (\(V_{int}\)) and are plotted in Fig. \ref{Fig Eyes} (circles) as a function of the percentage of heated area of the strip.
While the upper curve corresponds to heating the contacts side of the sample, the lower curve corresponds to heating the opposite side.

The temperature time dependence at the laser spot for these simulations is:

\begin{equation}
\frac{T-T_{0}}{T_{i}}= A \frac{(t/C)^N}{1+\frac{N}{M-N}(t/C)^M}
\label{Ttfunc}
\end{equation}
where \(T_0\) and \(T_i\) are the initial and the irreversibility temperatures, and \(t\) is the time elapsed since the laser pulse is triggered. 
The parameter values are \(N=4\), \(C=15\)~ns, \(M=4.3\) and \(A=2.8\). 
The essential feature of this curve is the asymmetry between the fast heating and the slow cooling of the illuminated area.
This behavior has been observed in simulations of heat diffusion for realistic values of diffusion coefficients of sample and substrate, and sample thickness \cite{SimTem}.
The particular values of the coefficients in Eq. (\ref{Ttfunc}) follow from a global fit to the experimental data. 
Observe that in our simulations the temperature rises from the initial value \(T_{0}=73.2\)~K to  \(T_{max}=90\)~K, which is higher than the irreversibility temperature \(T_{i}=86\)~K.

From the numerical solution of the evolution equation (Eq. (\ref{maxwell})) with the constitutive relation already discussed we have reconstructed the time evolution of the sheet current and the magnetic and electric fields inside the sample. 
Also, assuming the Josephson relation \(E=-v\times B\), we have obtained the mean velocity field of the vortices within the sample.

\begin{figure}
\centerline{\epsfig{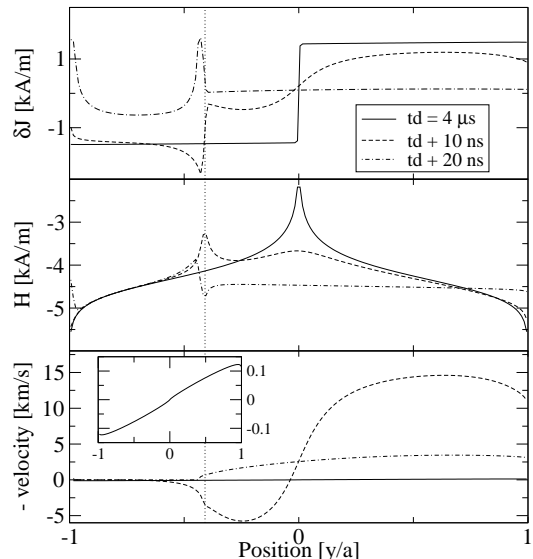}}
\caption{Space distribution of current density (upper), magnetic field (middle) and vortices velocities (lower) for different times since the start of the laser pulse. The inset in the lower panel details the symmetry of the initial condition.}
\label{Figdis}
\end{figure}

In Fig. \ref{Figdis} we show the calculated sheet current distribution (upper panel) at different times since the triggering of the laser pulse. 
The vertical line shows the border between cold (left) and heated (right) areas. 
We also show the magnetic field (middle panel) and the velocity field (lower panel) at the corresponding times. 
Observe that a positive sign in this last plot represents left moving vortices, while the applied field points in the negative \(x\) direction. 
Note that velocities after heating reach values of order \(10^4\)~m/s and  before heating \(10^2\)~m/s. 
To appreciate the different scales we have inserted in the lower figure the plot of the initial velocities with a higher resolution.
At \(t_d\) the velocities are odd because the applied field is homogeneous.
The calculated mean velocities are consistent with earlier observations of propagation of flux fronts\cite{velocities}.

The rearrangement of the vortices is triggered by the fast decay in the critical current produced by heating.
In fact, the critical sheet current drops to zero, since the temperature goes over the irreversibility line. 
While magnetic field penetration is easier from the right edge of the strip, some vortices enter the illuminated area from the left, meaning that we must have vortex motion in the cold area as well. 

Observe that a possible temperature dependence in the creep exponent \(n\) would not affect our results, since in the cold zone temperature is essentially constant, and in the hot zone the sheet current is over the critical one and so in the flux flow regime.
 
Our simulations also display two slower phenomena, not shown in Fig. \ref{Figdis}.
These are the relaxation towards an homogeneous distribution of vortices in the hot zone, with a characteristic time of the order of the \(\mu\)s, and then the change in magnetization in response to the changing applied field, which is even slower.

In conclusion, we have developed a non-conventional technique whereby an abrupt change in magnetic field on the sample is accomplished not by varying the applied field, but by heating the sample with a laser pulse. 
By scanning the laser pulse across the sample, delaying the application of the pulse with respect to the application of the external magnetic field, and by heating the sample both on and across the contacts' side, we have been able to observe different aspects of vortex dynamics on nanosecond time scales, and to discriminate between the different contributions to the measured signal. 
By means of numerical simulations, we have been able to reconstruct the evolution of the current and vortex motion within the sample in these fast time scales, thus opening a new time window for the study of vortex dynamics in high \(T_c\) superconductors.

We thank S. O. Valenzuela for fruitful discussions.
V.Bekeris and E.Calzetta are members of CONICET. 
This research was partially supported by 
UBACyT TX-90, 
UBACyT TW-12, 
CONICET 
PID \(N^{\circ }\) 4634 
\& 
PIP \(N^{\circ }\) 4452, 
Fundaci\'{o}n Sauber\'{a}n 
and 
Fundaci\'{o}n Antorchas.

\end{document}